\documentclass[conference]{IEEEtran}
\IEEEoverridecommandlockouts
\usepackage{cite}
\usepackage{amsmath,amssymb,amsfonts}
\usepackage{algorithmic}
\usepackage{graphicx}
\usepackage{multirow}
\usepackage{textcomp}
\usepackage{xcolor}
\usepackage{threeparttable}
\usepackage{booktabs}
\usepackage{graphicx}
\def\BibTeX{{\rm B\kern-.05em{\sc i\kern-.025em b}\kern-.08em
    T\kern-.1667em\lower.7ex\hbox{E}\kern-.125emX}}
\begin{document}

\title{Joint ASR and Speaker Role Tagging with Serialized Output Training\\
\thanks{We acknowledge \textit{Simons Foundation} for the funding.}
}
% \author{\IEEEauthorblockN{Author anonymized}
% \IEEEauthorblockA{\textit{dept. name of organization  anonymized} \\
% \textit{name of organization  anonymized}\\
% City, Country anonymized \\
% email address or ORCID anonymized}
\author{\IEEEauthorblockN{Anfeng Xu}
\IEEEauthorblockA{\textit{Viterbi School of Engineering} \\
\textit{University of Southern California}\\
Los Angeles, USA }
\and
\IEEEauthorblockN{Tiantian Feng}
\IEEEauthorblockA{\textit{Viterbi School of Engineering} \\
\textit{University of Southern California}\\
Los Angeles, USA }
\and
\IEEEauthorblockN{Shrikanth Narayanan}
\IEEEauthorblockA{\textit{Viterbi School of Engineering} \\
\textit{University of Southern California}\\
Los Angeles, USA }

}

\maketitle

\begin{abstract}
Automatic Speech Recognition systems have made significant progress with large-scale pre-trained models. However, most current systems focus solely on transcribing the speech without identifying speaker roles, a function that is critical for conversational AI. In this work, we investigate the use of serialized output training (SOT) for joint ASR and speaker role tagging. By augmenting Whisper with role-specific tokens and fine-tuning it with SOT, we enable the model to generate role-aware transcriptions in a single decoding pass. We compare the SOT approach against a self-supervised previous baseline method on two real-world conversational datasets. Our findings show that this approach achieves more than 10\% reduction in multi-talker WER, demonstrating its feasibility as a unified model for speaker-role aware speech transcription.
\end{abstract}


\section{Introduction}
\begin{figure}[t]
  \centering
  \includegraphics[width=\linewidth]{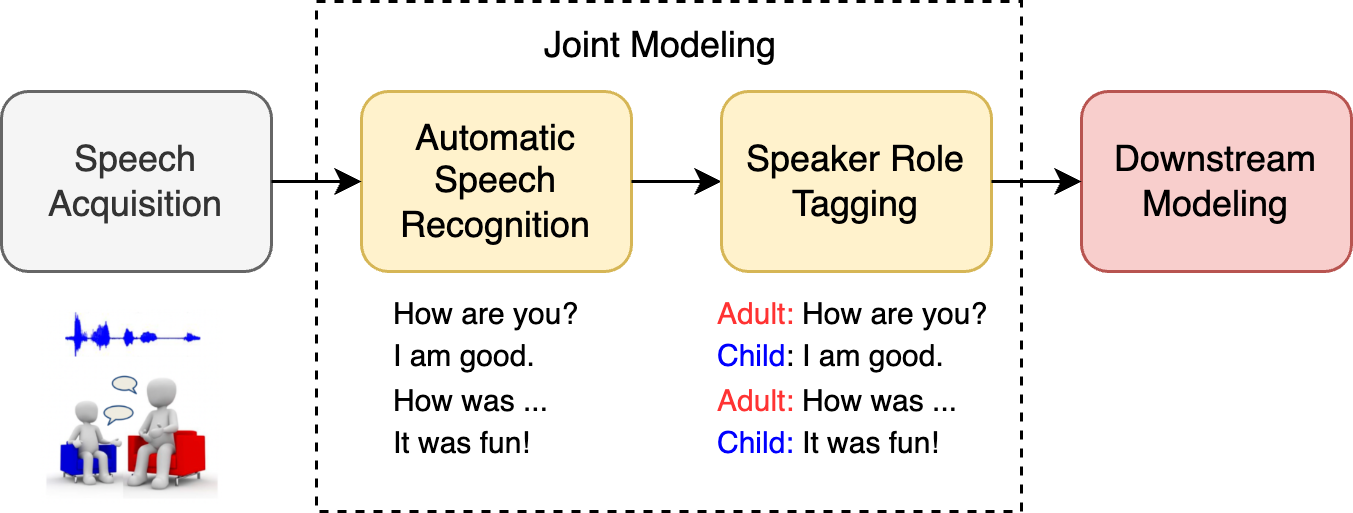}
  \caption{Overview of the spoken conversation analysis pipeline. The pipeline begins with speech acquisition, followed by the proposed \textbf{joint modeling} that integrates (i) automatic speech recognition (ASR) to transcribe spoken utterances and (ii) speaker role tagging to assign each utterance to a speaker (e.g., Adult or Child). The processed transcript, enriched with speaker role information, is then used in downstream modeling tasks.}
  \label{figures:pipeline}
\end{figure}

Automatic Speech Recognition (ASR) systems have achieved remarkable progress in recent years, driven by large-scale pre-trained models capable of transcribing speech across diverse languages and domains \cite{prabhavalkar2023end}. Likewise, recent advancements in speaker diarization have improved the accuracy and robustness of automatic detection of ``who spoke when'' \cite{park2022review}. Despite the advances, these systems do not identify or leverage the conversational \textit{speaker role} (e.g., child-parent, doctor-patient), which could enrich the ASR service by providing relevant context information. Specifically, this limitation reduces the interpretability of ASR service for applications in computational behavioral analysis with distinct speaker roles, such as child-adult interactions, doctor-patient conversations, and pilot-controller communications. In such scenarios, knowing not just what was said but who said it, and \textit{in what role}, is critical for downstream processing and understanding of the interaction. Thus, we aim to transcribe speech while simultaneously tagging the corresponding speaker roles, as illustrated in Figure~\ref{figures:pipeline}.

Traditionally, ASR and speaker role tagging (SRT) have been treated as separate tasks, often implemented in a pipeline where speech is first transcribed and then passed to a secondary system for speaker attribution or role classification \cite{blatt2024joint}. Such multi-stage processing approaches are prone to error propagation and do not effectively leverage the relation between the speaker's role and their communication styles, such as word choice and sentence structure.
% to leverage potential synergies between the two tasks. 
Moreover, existing joint speaker diarization and ASR approaches \cite{shafey2019joint, cornell2024one, park2024sortformer} focus on speaker identity rather than the speaker role information, limiting their applicability to scenarios where speaker roles carry semantic significance beyond mere speaker distinction.

Serialized output training (SOT) has emerged as a promising end-to-end framework to jointly model lexical content and speaker information by representing both in a single serialized output sequence \cite{kanda2020serialized}. Prior work has shown the effectiveness of SOT in multi-speaker ASR and speaker diarization tasks, often using autoregressive models with attention-based encoder-decoder (AED) architectures \cite{cornell2024one, park2024sortformer}. However, its potential for speaker role tagging, where speaker tokens encode meaningful conversational roles rather than arbitrary speaker IDs, remains underexplored.

In this work, we explore Whisper \cite{radford2023robust}, a large-scale speech foundation model based on an attention-based encoder-decoder (AED) architecture trained on 680k hours of multilingual and multitask audio, for joint ASR and SRT within the SOT framework. The AED architecture with language modeling and large-scale pre-training enable Whisper's strong speech and lexical representation capabilities \cite{xu2025large, wang2023whislu}, making it well-suited for incorporating speaker role information, which often manifests through distinct acoustic and semantic patterns. We extend its vocabulary with special tokens for speaker roles and fine-tune the model on two conversational speech datasets, Playlogue and MMCSG, which include annotations for child-adult and self-other speaker roles, respectively. Our contributions are summarized as follows:

\begin{itemize}
    \item Using the SOT training framework with role-aware speaker tokens, we propose a Whisper-based solution for joint ASR and SRT. To the best of our knowledge, this is one of the earliest efforts exploring SOT using Whisper. 
    \item We benchmark our approach against a self-supervised learning (SSL) based baseline, showing $10\% \sim 15\%$ reduction in terms of multi-talker WER.
    \item We experiment on two real-world conversational speech datasets, evaluating the impact of model size, conversational context conditioning, and audio window length on the ASR and speaker role tagging performance.
    \item Our findings demonstrate the feasibility of using SOT for producing role-aware transcriptions in conversational settings, opening new opportunities for speaker-aware speech technologies.
\end{itemize}

\section{Background}

\subsection{Serialized Output Training}
Serialized output training (SOT) \cite{kanda2020serialized} has emerged as an effective formulation for unifying multi-speaker recognition and transcription tasks. Originally proposed for end-to-end overlapped speech recognition, SOT reformulates the target sequence to include both lexical tokens and explicit \textit{speaker} (or \textit{speaker change}) tags in a single serialized target sequence. Consequently, SOT allows sequence-to-sequence models to learn the content and speaker attribution jointly using AED architectures, eliminating the need for separate diarization or source separation modules typically used in pipeline systems. Specifically, the multi-headed attention mechanism \cite{vaswani2017attention} enables the model to focus on different speakers within a single AED architecture without needing multiple encoders or heads per speaker. This approach has since been explored in various multi-talker ASR tasks \cite{lin2023directional, lin2024agadir, feng2025directional}. Additionally, using the SOT training framework, \cite{cornell2024one} has proposed a unified model for joint ASR and speaker diarization, while \cite{kanda2022streaming} has provided an effective streaming multi-talker ASR solution. In our work, we adopt this SOT framework to speaker role tagging, where the goal is not only to distinguish between speakers but to assign meaningful speaker role labels (e.g., \textless child\textgreater, \textless adult\textgreater), enabling role-aware transcription in conversational settings.

\subsection{Joint ASR and Speaker Role Tagging}
\begin{figure*}[!t]
  \centering
  \includegraphics[width=0.9\textwidth]{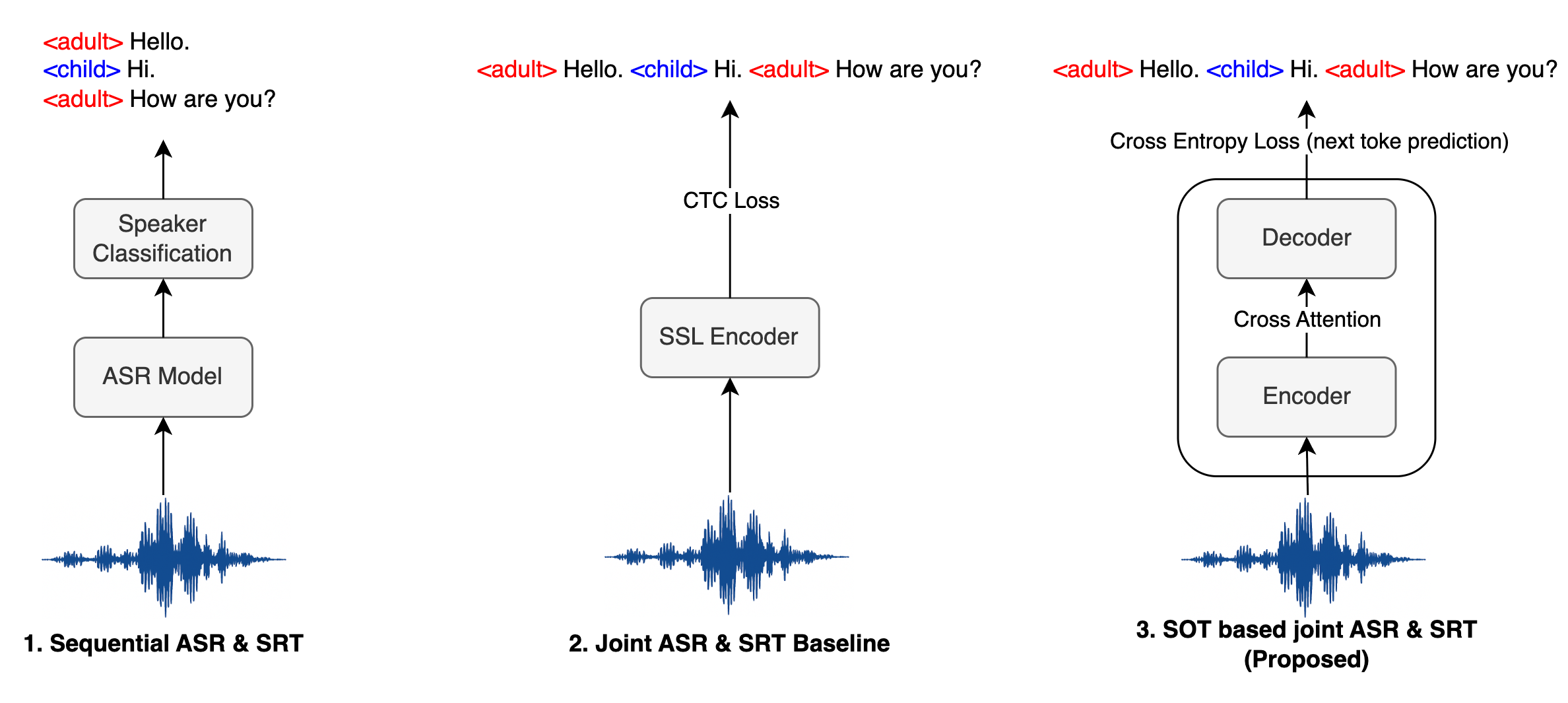}
  \caption{Overview of the modeling pipelines illustrating (1) a traditional sequential approach, (2) a joint baseline approach, and (3) the proposed SOT-based approach with Whisper. SRT means Speaker Role Tagging, and SSL means a model based on Self-supervised learning.}
  \label{figures:models}
\end{figure*}

The modeling pipeline solutions are illustrated in Figure~\ref{figures:models}. Traditionally, ASR and Speaker Role Tagging (SRT) are treated as two distinct tasks. For example, \cite{zuluaga2023bertraffic} uses ASR followed by BERT\cite{devlin2019bert} based pilot and air-traffic controller speaker tagging. Furthermore, \cite{sun2025said} aligns separated results from Whisper-based ASR and wav2vec 2.0 \cite{baevski2020wav2vec} based child-adult speaker classification for speaker-tagged transcripts. Recently, \cite{blatt2024joint} has explored jointly learning ASR and air-traffic controller speaker tags with the serialized target output format similar to SOT, outperforming the sequential modeling approach. In doing so, they have used fine-tuned self-supervised models such as wav2vec 2.0 with CTC loss. However, even though AED-based approaches are promising and suitable for the joint ASR and SRT given the success with SOT, to the best of our knowledge, little work has explored their use for this task. The Whisper model is an ideal candidate for this task, as it is an AED-based speech foundation model that has shown strong performance in ASR and related tasks such as speech translation. In addition, prior works have shown that the Whisper encoder contains rich representations of speaker role information (e.g., child-adult) \cite{xu2024exploring, xu2025data, feng2025vox}. 

\begin{figure}[t]
  \centering
  \includegraphics[width=1\linewidth]{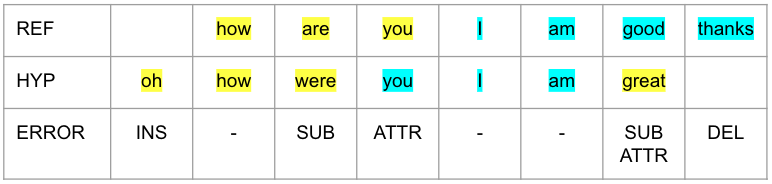}
  \caption{Error components for mtWER. The yellow and blue colors highlight two separate speaker roles. REF is the ground-truth transcript, while HYP is the inferred hypothesis.}
  \label{figures:mtWER}
\end{figure}

\section{Method}
\subsection{Problem Setup}

\subsubsection{Problem Formulation.}
Given an input audio $\mathbf{x}$, our goal is to produce sequences $\mathbf{y} = (y_1, y_2, \ldots, y_T)$ and $\mathbf{s} = (s_1, s_2, \ldots, s_T)$ where the token $y_t$ belongs to a word and the token $s_t$ indicates the corresponding speaker role token. The tokens are sorted by the utterance-level timestamps.

\subsubsection{Evaluation Metrics.}
We employ \textbf{multi-talker Word Error Rate (mtWER)}, which is originally used in the CHiME-8 MMCSG Challenge\cite{zmolikova2024chime}. It extends the conventional WER metric by accounting for both transcription accuracy and speaker attribution consistency in multi-speaker scenarios. Specifically, mtWER computes speaker-attribution errors as well as the word errors (substitutions, insertions, deletions). This makes mtWER a stricter and more realistic measure for joint ASR and speaker role tagging tasks, as it penalizes both recognition and attribution mistakes. We report mtWER for each speaker role individually, for a more comprehensive understanding. An example with all the error components is shown in Figure~\ref{figures:mtWER}. Mathematically, mtWER is calculated as:
$$
mtWER_{s} = \frac{INS_{s}+DEL_{s}+SUB_{s}+ATTR_{s}}{NREF_{s}},
$$
where $s$ refers to the specific speaker role (e.g., child). For each speaker role $s$,  $INS_{s}, DEL_{s}, SUB_{s}$, and $ATTR_{s}$ refer to insertion, deletion, substitution, speaker attribution errors, respectively, while  $NREF_{s}$ refers to the total number of reference words from $s$. The speaker attribution error occurs when a word from $s$ is assigned to a different speaker role, including when the word is substituted. 

In addition to mtWER, we also report \textbf{WER} and \textbf{Attribution Error Rate (AER)} per speaker role, defined as:
$$
WER_{s} = \frac{INS_{s}+DEL_{s}+SUB_{s}}{NREF_{s}},
$$
$$
AER_{s} = \frac{ATTR_{s}}{NREF_{s}}.
$$

\subsection{SOT-based joint ASR and SRT using Whisper}
% As an attention-based encoder-decoder (AED) model, Whisper generates output in an autoregressive manner, making it naturally suited for serialized output training (SOT). Unlike CTC-based models, Whisper can condition on previously generated tokens, allowing it to model dependencies across both lexical and speaker role tokens in the serialized sequence. 

We fine-tune Whisper models with a modified vocabulary that includes special speaker role tokens. Specifically, we add the speaker tokens ``\textless spk0\textgreater'' and ``\textless spk1\textgreater'' to the Whisper tokenizer vocabulary as special tokens. In the preprocessing and postprocessing steps, the speaker roles (e.g., child-adult) are mapped to ``\textless spk0\textgreater'' and ``\textless spk1\textgreater''. During both training and inference, the decoder initializes decoding with the following special tokens: \textless startoftranscript\textgreater\textless en\textgreater\textless transcribe\textgreater\textless notimestamps\textgreater.

We evaluate different Whisper model sizes, including tiny, small, and large, to understand the trade-offs between model capacity and role tagging performance. In addition, we explore both the English-only and multilingual pre-trained variants of Whisper, along with other modeling decisions, including audio input length and encoder freezing during training. We also examine the performance when conditioning on the preceding transcript as additional context, motivated by Whisper’s ability to leverage history through its autoregressive decoding. This setup allows us to assess whether conversational context improves role consistency in multi-turn interactions. 

\subsection{Baseline}
For the baseline comparison, we use the approach in \cite{blatt2024joint} that uses SSL-based models fine-tuned with CTC-loss for joint ASR and SRT predictions. To the best of our knowledge, this is the current state of the art (SOTA) published approach for this problem setup. While the original work uses wav2vec 2.0 \cite{baevski2020wav2vec} and XLS-R \cite{babu2021xls}, we choose the WavLM-large model \cite{chen2022wavlm}, as it is the current SOTA model for the SUPERB speech SSL modeling benchmark \cite{yang2021superb}.

\section{Experiments}
\subsection{Datasets}
We use two publicly available conversational speech datasets, Playlogue \cite{kalanadhabhatta2024playlogue} and MMCSSG \cite{zmolikova2024chime}, that contain annotated transcriptions with speaker role tags and speaker timestamp information.

\subsubsection{Playlogue}
The Playlogue dataset \cite{kalanadhabhatta2024playlogue} consists of over 33 hours of long-form, naturalistic adult-child conversations collected from the TalkBank system \cite{macwhinney2007talkbank}. This dataset spans three play-based corpora and one narrative corpus involving preschool-aged children (3–5 years old). The training, validation, and test sets contain approximately $19h$, $6h$, and $8h$ of audio data, respectively. These recordings capture spontaneous interactions during free-play sessions between children and adults, including parents, examiners, and researchers. We treat the child-adult distinction as the target speaker roles. The dataset includes audio recordings with word-aligned transcripts with child-adult speaker roles, produced via forced alignment using NVIDIA NeMo \cite{kuchaiev2019nemo}. The dataset presents a unique challenge due to the variability in child speech, child-directed adult speech, and interaction dynamics, making it an ideal testbed for evaluating joint ASR and SRT systems. To prepare the SOT target output, we define each utterance based on the official transcriptions. The mean child and adult utterance durations are 1.72s and 1.61s, respectively.

\subsubsection{MMCSG}
The Multi-Modal Conversational Speech Grounding (MMCSG) dataset, released as part of the CHiME-8 Challenge Task 3 \cite{zmolikova2024chime}, consists of multi-modal recordings of dyadic conversations collected using Meta’s Project Aria glasses. The dataset includes 172 training (8.5h), 169 validation (8.4h), and 189 test (9.4h) sessions, each featuring spontaneous conversational speech between two participants recorded in natural acoustic environments. Each participant wore a pair of Aria glasses that contained a 7-channel spatial microphone array. In this study, we only use the 0th channel as the mono-channel input to the speech models. For each session, the dataset provides human-annotated word-level transcriptions with speaker labels distinguishing between the glasses wearer (self) and their conversational partner (other). We treat the self-other distinction as the target speaker roles. As the original annotation provides per-word transcription, we have prepared the utterance-by-utterance transcription for SOT training framework by treating adjacent words from the same speaker within 0.3s as a single utterance. The mean self and other utterance durations are  2.34s and 2.18s, respectively.

\subsection{Experimental Setup}
We use the official train, validation, and test splits for both the Playlogue and MMCSG datasets. Using ground-truth silence boundaries, we chunk the audio into segments that are as long as possible while remaining under the 15-second maximum duration, resulting in around 12.5 seconds average durations. The results from all sessions are aggregated to calculate the final evaluation metrics. We use a single NVIDIA RTX A6000 48GB GPU for all the experiments. 

For Whisper models, we apply the Whisper English text normalizer to the texts as a pre-processing step. We train the models for 1500 steps with a batch size of 32. Adam optimizer is used with the learning rates of $1e-5$, $2e-6$, and $1e-6$ for Whisper-tiny, Whisper-small, and Whisper-large models, respectively. We choose the model with the lowest validation loss calculated at every 100 steps. The validation loss is selected in favor of validation WER due to computational cost. 

When fine-tuning the baseline WavLM model, the numbers from the Whisper English text normalizer (e.g., 1st, 5th, 12) are mapped to textual numbers (e.g., first, fifth, twelve). We initialize the model from the WavLM checkpoint (\textit{patrickvonplaten/wavlm-libri-clean-100h-large}) on Hugging Face \cite{wolf2019huggingface}, which was fine-tuned for ASR using 100 hours of clean LibriSpeech data. We train for 4000 steps with Adam optimizer, batch size of 32 and learning rate of 3e-4. We choose the model with the lowest WER, with speaker role tokens treated as a word, evaluated every 200 steps.

\section{Results}

\begin{table}[t]
\footnotesize

  \caption{Results using the Playlogue dataset. The numbers are shown in percentage($\%$).}
  \label{tab:results-playlogue}
  \centering
  \begin{tabular*}{0.8\linewidth}{l c c c c}
    \toprule
    \multicolumn{1}{l}{\textbf{Model}} & \textbf{speaker} & \textbf{mtWER} & \textbf{WER} & \textbf{AER} \\
    \cmidrule(lr){1-1} \cmidrule(lr){2-2} \cmidrule(lr){3-5} 
    WavLM-large & child & $69.3$ & $63.1$ & $6.2$ \\
    (baseline) & adult  & $37.0$ & $31.8$ & $5.2$ \\
    \cmidrule(lr){1-1} \cmidrule(lr){2-2} \cmidrule(lr){3-5} 
    \multirow{2}{*}{Whisper-tiny} & child & $127.5$ & $119.5$ & $8.0$ \\
    & adult  & $67.8$ & $55.1$ & $12.7$ \\
    \cmidrule(lr){1-1} \cmidrule(lr){2-2} \cmidrule(lr){3-5} 
    \multirow{2}{*}{Whisper-small} & child & $57.8$ & $53.0$ & $4.8$ \\
    & adult  & $24.1$ & $22.5$ & $\mathbf{1.6}$ \\
    \cmidrule(lr){1-1} \cmidrule(lr){2-2} \cmidrule(lr){3-5} 
    \multirow{2}{*}{Whisper-large} & child & $\mathbf{54.8}$ & $\mathbf{51.2}$ & $\mathbf{3.6}$ \\
    & adult & $\mathbf{23.6}$ & $\mathbf{20.5}$ & $3.1$ \\
                          
    \bottomrule
  \end{tabular*}
\end{table}

\begin{table}[t]
% \footnotesize

  \caption{Results using the MMCSG dataset. The numbers are shown in percentage($\%$).}
  \label{tab:results-mmcsg}
  \centering
  \begin{tabular*}{0.8\linewidth}{l c c c c}
    \toprule
    \multicolumn{1}{l}{\textbf{Model}} & \textbf{speaker} & \textbf{mtWER} & \textbf{WER} & \textbf{AER} \\
    \cmidrule(lr){1-1} \cmidrule(lr){2-2} \cmidrule(lr){3-5} 
    WavLM-large & self  & $26.7$ & $20.9$ & $5.8$ \\
    (baseline) & other & $34.0$ & $31.3$ & $\mathbf{2.7}$ \\
    \cmidrule(lr){1-1} \cmidrule(lr){2-2} \cmidrule(lr){3-5} 
    \multirow{2}{*}{Whisper-tiny} & self  & $58.9$ & $27.8$ & $31.1$ \\
                          & other & $54.3$ & $50.9$ & $3.4$ \\
    \cmidrule(lr){1-1} \cmidrule(lr){2-2} \cmidrule(lr){3-5} 
    \multirow{2}{*}{Whisper-small} & self  & $\mathbf{16.3}$ & $\mathbf{10.8}$ & $\mathbf{5.5}$ \\
    & other & $\mathbf{23.1}$ & $\mathbf{19.4}$ & $3.7$ \\
    \cmidrule(lr){1-1} \cmidrule(lr){2-2} \cmidrule(lr){3-5} 
    \multirow{2}{*}{Whisper-large} & self  & $23.2$ & $12.7$ & $10.5$ \\
                          & other & $26.6$ & $21.5$ & $5.1$ \\
    \bottomrule
  \end{tabular*}
\end{table}

Tables~\ref{tab:results-playlogue} and \ref{tab:results-mmcsg} show the ASR and speaker tagging results from the WavLM baseline and Whisper models using Playlogue and MMCSG datasets. The WavLM-large has $ 317 M$ parameters, while the Whisper models have $ 39 M$, $ 244 M$, and $ 1550 M$ parameters, respectively, for the tiny, small, and large variants. Overall, Whisper-large performs the best for the Playlogue dataset, while Whisper-small performs the best for the MMCSG dataset. The mtWER scores are substantially reduced by around $10\%$ to $15\%$ across the speaker roles and datasets compared to the WavLM-large baseline. As expected, the WERs for the ``child'' and ``other'' speaker roles are higher than those for their respective counterparts. We also note that the low AER using the Whisper models, especially for the Playlogue dataset, indicate their capacity to handle speaker differentiation in conversational speech. 

We see that the Whisper-tiny model fails to perform well, likely due to its small parameter size, which limits its capacity to capture the nuanced acoustic and speaker-role information required for effective joint ASR and role tagging performance. In addition, we reason that Whisper-small outperforms Whisper-large on the MMCSG dataset because the limited size of the training data ($ 8.5 h$) constrains the effective adaptation of larger models, whose stronger pretrained priors may be less amenable to fine-tuning under low-resource conditions.

We have also experimented with the English-only versions of the Whisper-tiny and Whisper-small models. As shown in Table~\ref{tab:english-only}, the English-only variants generally outperform their multilingual counterparts. We attribute this to the fact that, when adapting Whisper to a new task involving additional tokens, initializing from a model pretrained on the same language reduces the mismatch between pretraining and fine-tuning objectives, facilitating more effective transfer learning. For the rest of the paper, we use Whisper-small.en for further experiments and analysis.

\begin{table}[t]
\footnotesize

  \caption{Whisper models with English only pertaining. $\Delta$ mtWER denotes the change from the multilingual version.}
  \label{tab:english-only}
  \centering
  \begin{tabular*}{0.93\linewidth}{l c c c c}
    \toprule
    \multicolumn{1}{l}{\textbf{Model}} & \textbf{Dataset} & \textbf{speaker} & \textbf{mtWER} & \textbf{$\Delta$ mtWER} \\
    \cmidrule(lr){1-1} \cmidrule(lr){2-2} \cmidrule(lr){3-5} 
    \multirow{2}{*}{Whisper-tiny.en} & \multirow{2}{*}{Playlogue}  & child & $116.4$ & $-11.1$ \\
    &  & adult & $44.0$ & $-23.8$ \\
    \cmidrule(lr){1-1} \cmidrule(lr){2-2} \cmidrule(lr){3-5} 
    \cmidrule(lr){1-1} \cmidrule(lr){2-2} \cmidrule(lr){3-5} 
    \multirow{2}{*}{Whisper-small.en} & \multirow{2}{*}{Playlogue}  & child & $60.6$ & $+2.8$ \\
    &  & adult & $22.2$ & $-1.9$ \\
    \cmidrule(lr){1-1} \cmidrule(lr){2-2} \cmidrule(lr){3-5} 
    \multirow{2}{*}{Whisper-tiny.en} & \multirow{2}{*}{MMCSG}  & self & $24.6$ & $-34.3$ \\
    &  & other & $44.6$ & $-9.7$ \\
    \cmidrule(lr){1-1} \cmidrule(lr){2-2} \cmidrule(lr){3-5} 
    \cmidrule(lr){1-1} \cmidrule(lr){2-2} \cmidrule(lr){3-5} 
    \multirow{2}{*}{Whisper-small.en} & \multirow{2}{*}{MMCSG}  & self & $15.9$ & $-0.4$ \\
    &  & other & $20.2$ & $-2.9$ \\
    \bottomrule
  \end{tabular*}
\end{table}

\section{Analysis}

\subsection{Does a longer input audio length help?}
\begin{figure}[t]
  \centering
  \includegraphics[width=0.9\linewidth]{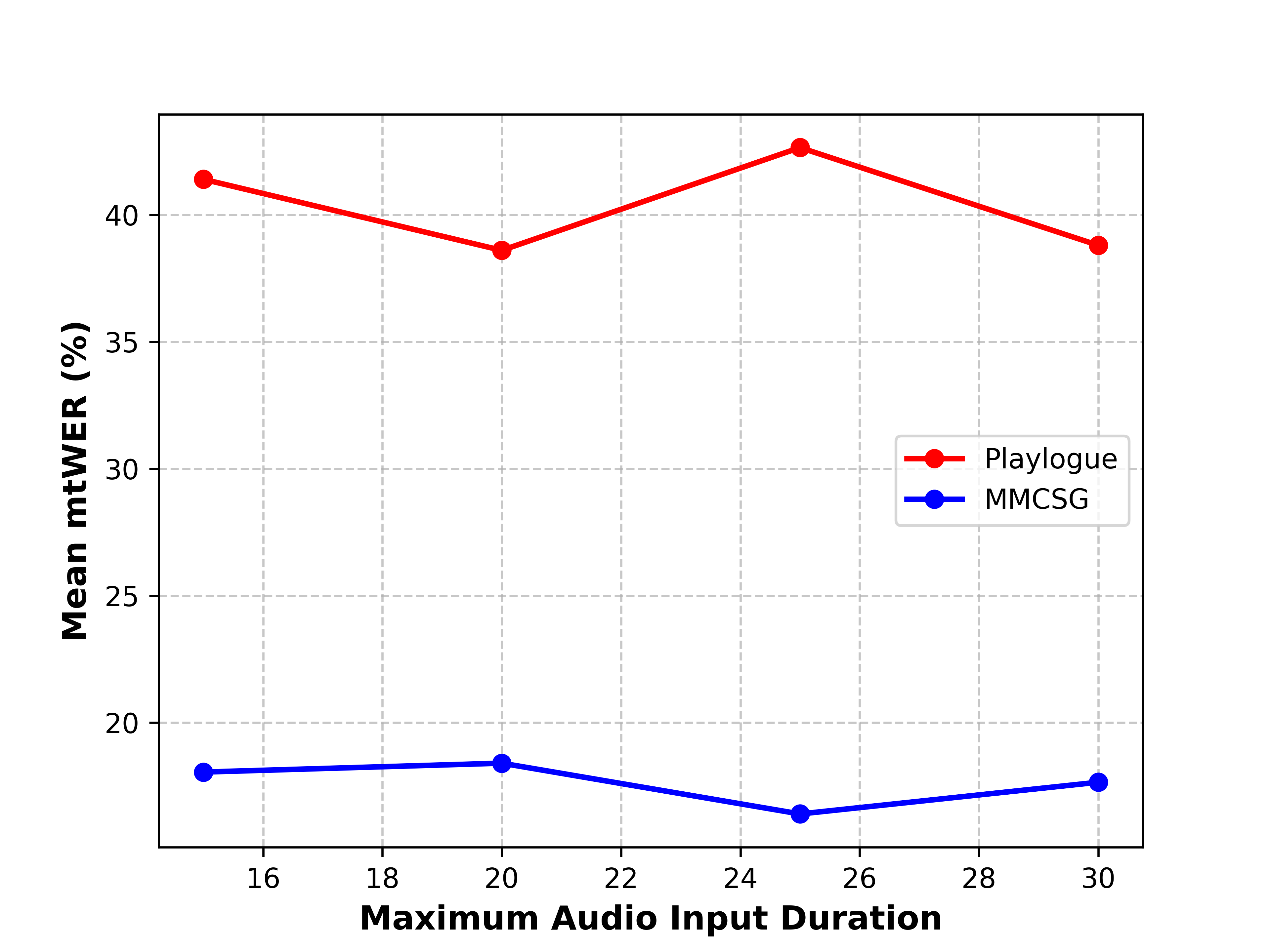}
  \caption{mtWERs when varying audio input duration using Whisper-small.en.}
  \label{figures:duration}
\end{figure}

Although we have chosen the maximum audio input duration to 15s, which is a similar length to the baseline \cite{blatt2024joint}, Whisper allows for a maximum of 30s audio input. Longer audio input provides more context for the Whisper models, but the longer decoding process may promote more error propagation and hallucination. To explore whether longer audio input helps improve the joint ASR and SRT, we have experimenting by varying the audio input lengths across 15s, 20s, 25s, and 30s. Figure~\ref{figures:duration} reports the mean mtWER of 2 speakers for each dataset. Longer input audio length does not show substantial improvements in resulting mtWER. Thus, we keep the same 15s input audio length for the rest of the paper.

\subsection{Impact of freezing the encoder for joint ASR and SRT}
To assess the impact of encoder fine-tuning on speaker role-aware ASR performance, we have frozen the encoder layers of the Whisper model during training. Table~\ref{tab:freeze-encoder} presents results on both the Playlogue and MMCSG datasets when the encoder is kept frozen. Overall, freezing the encoder leads to performance degradation in most conditions. For the Playlogue dataset, we observe a substantial increase in mtWER for both child and adult speaker roles, with worse degradation for the ``child'' speech. 
For the MMCSG dataset, we see a substantial increase in mtWER only for the ``other'' speech. These findings emphasize that encoder fine-tuning is generally beneficial for improving role-aware transcription accuracy, particularly when dealing with acoustically mismatched or highly variable speech such as ``child'' or ``other'' speech.

\subsection{Does preceding transcript conditioning help?}
For audio inputs longer than 30 seconds, the original Whisper implementation prepends the predicted tokens from the last segment before the start of the transcription for a final continuous transcript. For the joint ASR and speaker role tagging, we hypothesize that having preceding conversations may help Whisper understand the conversational context and improve the ASR inference. To explore this, we have experimented with a training setup in which tokens from the preceding transcription are prepended before the token \textless startoftranscript\textgreater. At inference time, the model receives the predicted tokens from the preceding segment.

Contrary to our hypothesis, Table~\ref{tab:context} shows that the joint ASR and SRT performance degraded with the preceding transcript conditioning. A possible explanation is the discrepancy between training and inference: while ground-truth preceding text tokens are used during training, the model relies on its own predicted tokens during inference. This mismatch may be exacerbated in challenging datasets, leading to error propagation and reduced overall performance.

\begin{table}[t]
\footnotesize
  \caption{Results when encoders are frozen during training. $\Delta$ mtWER denotes the change compared to not freezing the encoders during training.}
  \label{tab:freeze-encoder}
  \centering
  \begin{tabular*}{0.7\linewidth}{l c c c}
    \toprule
    \multicolumn{1}{l}{\textbf{Dataset}} & \textbf{speaker} & \textbf{mtWER} & \textbf{$\Delta$ mtWER} \\
    \cmidrule(lr){1-1} \cmidrule(lr){2-4} 
    \multirow{2}{*}{Playlogue}  & child & $71.5$ & $+10.9$ \\
    & adult & $27.0$ & $+4.8$ \\
    \cmidrule(lr){1-1} \cmidrule(lr){2-4} 
    \multirow{2}{*}{MMCSG}  & self & $15.6$ & $-0.3$ \\
    & other & $40.8$ & $+20.5$ \\
    \bottomrule
  \end{tabular*}
\end{table}

\begin{table}[t]
\footnotesize
  \caption{Results with preceding transcript conditioning. $\Delta$ mtWER denotes the change compared to not using preceding transcript conditioning.}
  \label{tab:context}
  \centering
  \begin{tabular*}{0.7\linewidth}{l c c c}
    \toprule
    \multicolumn{1}{l}{\textbf{Dataset}} & \textbf{speaker} & \textbf{mtWER} & \textbf{$\Delta$ mtWER} \\
    \cmidrule(lr){1-1} \cmidrule(lr){2-2} \cmidrule(lr){3-4} 
    \multirow{2}{*}{Playlogue}  & child & $67.6$ & $+7.0$ \\
    & adult & $27.7$ & $+5.5$ \\
    \cmidrule(lr){1-1} \cmidrule(lr){2-2} \cmidrule(lr){3-4} 
    \multirow{2}{*}{MMCSG}  & self & $20.4$ & $+4.5$ \\
    & other & $28.8$ & $+8.6$ \\
    \bottomrule
  \end{tabular*}
\end{table}

\section{Conclusion}
This paper presents a joint approach for ASR and speaker role tagging using SOT. By fine-tuning Whisper for SOT with role-specific tokens on real-world conversational datasets, we enable it to generate transcriptions with integrated speaker role labels. Experiments show this approach outperforms an SSL-based baseline. Further analysis illustrates how model size, language pretraining, encoder freezing, and context usage affect the performance. Our findings highlight Whisper's potential as a strong foundation for speaker-aware ASR under SOT, with future directions including improved contextual modeling and generalization to broader applications.
% This paper has explored the use of Whisper for joint automatic speech recognition and speaker role tagging through serialized output training. By incorporating role-specific tokens into Whisper’s tokenizer and fine-tuning on real-world conversational datasets, we have enabled the model to generate transcriptions with explicit speaker role labels directly integrated into the output sequence.

% Our experiments demonstrate that Whisper can effectively handle this joint task, outperforming the SSL-based baseline. We have analyzed the impact of model size, language pretraining, audio input length, freezing the encoder, and conversational context. Results indicate that model size should be matched to data scale, freezing the encoder degrades performance, English-only pretraining improves adaptation, and naive use of previous context can harm performance due to training-inference mismatch.

% These findings establish Whisper as a strong foundation for end-to-end speaker-aware ASR. Future work may include designing better contextual modeling,  leveraging multimodal cues, and investigating Whisper's capability for related tasks such as joint ASR and speaker diarization.

\bibliographystyle{IEEEtran}
\bibliography{refs}

\end{document}